\documentclass[letterpaper,12pt]{article}
\usepackage{graphicx}
\usepackage{color,amssymb,enumerate,float,amsmath}
\usepackage{tikz}
\usetikzlibrary{matrix}
\usepackage{caption}
\usepackage{subcaption}
\usepackage{cite}
\usepackage{ifpdf}
\ifpdf
	\usepackage[pdftex,unicode,implicit]{hyperref}

	\hypersetup{
  	pdftitle     = {}, 
  	pdfkeywords  = {},
  	pdfauthor    = {},
  	pdfcreator   = {pdf\LaTeXe\ with package \flqq hyperref\frqq},
  	pdfproducer  = {pdf\LaTeXe\ with package \flqq hyperref\frqq},
  	pdfpagemode  = UseNone,  
  	pdffitwindow = true,  
  	unicode      = true,
  	plainpages   = true,
  	colorlinks   = true,  
  	citecolor    = black,  
  	urlcolor     = blue, 
  	linkcolor    = black
	}

\else

  \usepackage[unicode,implicit]{hyperref}

\fi 

\makeatletter
\@addtoreset{equation}{section}
\makeatother

\begin{document}
	
\thispagestyle{empty}

\begin{center}
{\bf \LARGE 
Quantum corrections to the Casimir effect in a scalar Ho\v{r}ava-Lifshitz theory with rough plates at low temperature}
\vspace*{15mm}

{\large Claudio B\'orquez}$^{1,a}$
{\large and 
Byron Droguett}$^{2,b}$
\vspace{3ex}

$^1${\it Facultad de Ingenier\'ia, Universidad San Sebasti\'an, Lago Panguipulli 1390, Puerto Montt, Chile.
}

$^2${\it Department of Physics, Universidad de Antofagasta, 1240000 Antofagasta, Chile.}

\vspace{3ex}

$^a${\tt 
claudio.borquez@uss.cl},
$^b${\tt 
byron.droguett@uantof.cl
}

\hspace{.5em}

{\bf Abstract
}
\begin{quotation}{\small\noindent
}

We investigate the Casimir effect of a massive, self-interacting real scalar field with quartic coupling in a $(3+1)$-dimensional Ho\v{r}ava-Lifshitz-type theory, subject to rough boundaries obeying Dirichlet boundary conditions. Using the effective action and generalized $\zeta$-function approach, we derive the renormalized effective potential, incorporating anisotropic scaling, boundary roughness, and temperature in the low-temperature regime. We determine the topological mass at one-loop order and the Casimir energy up to two loops. In the massless limit, the topological mass and the two-loop Casimir contribution exhibit infrared divergences for odd values $z>1$, associated with the absence of an intrinsic mass scale, while the one-loop Casimir energy remains finite. For $z=1$, the standard relativistic Dirichlet result is recovered in the zero-temperature and vanishing-roughness limits.

\end{quotation}
\vspace{3ex}
\end{center}

\thispagestyle{empty}

\newpage


\section{Introduction}

The Casimir effect is one of the most remarkable manifestations of quantum vacuum fluctuations and provides a natural setting in which quantum field theory can be connected with measurable macroscopic phenomena. In his seminal work, Casimir showed that two parallel, neutral conducting plates in $(3+1)$-dimensional spacetime experience an attractive force arising from the modification of the electromagnetic vacuum fluctuations by the boundary conditions \cite{Casimir:1948dh}. The Casimir effect has been extensively investigated both theoretically and experimentally \cite{Lamoreaux:1996wh,Bressi:2002fr}. It has consequently become an important laboratory for probing quantum vacuum phenomena. Subsequent studies have established that the Casimir interaction is sensitive to the physical properties of the system, including boundary conditions, material properties, topology, temperature, and geometry \cite{Mazur,Bordag:2001qi,Teo:2011kt,Fosco:2011xx,Bimonte:2012dqc,Zhao:2006rr,Beneventano:2004zd,Mota:2023cmr,Bellucci:2019ybj}. External background can also modify vacuum fluctuations, as illustrated by the effects of magnetic fields \cite{Beneventano:2005sd,Erdas:2013jga,Erdas:2015yac,Droguett:2025frq} and gravitational fields \cite{Ford:1976fn,Nazari:2015oha,Nazari:2025bae,Muniz:2014dga,Borquez:2023cuf}. Geometry is also a key ingredient in determining the nature of the interaction. For parallel plates, the Casimir force is attractive, whereas a spherical shell can produce a repulsive force \cite{Boyer:1968uf}, and other configurations can cause lateral stresses \cite{Emig:2002qpp}. These results demonstrate that the spectrum of vacuum fluctuations, and hence the resulting Casimir interaction, is strongly dependent on the geometry and physical properties of the boundaries.

Modifications of spacetime symmetries provide a natural framework for investigating how quantum vacuum fluctuations are affected by departures from Lorentz invariance. In particular, Lorentz-violating field theories can modify the dispersion relation of quantum fields and consequently alter the spectrum responsible for the Casimir interaction \cite{deMello:2022tuv,Cruz:2017kfo,Erdas:2020ilo,Cruz:2018bqt,Erdas:2021xvv,Droguett:2024tpe}. A prominent realization is provided by the Ho\v{r}ava-Lifshitz theories, in which the temporal and spatial coordinates exhibit different scaling dimensions. The resulting anisotropic structure allows for higher spatial derivatives while keeping the time derivative at second order, providing a framework with improved ultraviolet behavior \cite{Horava:2009uw, Anselmi:2008bq}. The consequences of anisotropic scaling for the Casimir effect have been investigated in a variety of Ho\v{r}ava-Lifshitz models. These studies have considered scalar and fermionic fields, as well as the effects of external magnetic fields and finite temperature, showing that the modified dispersion relation can significantly affect the vacuum energy induced by the boundaries \cite{Ferrari:2010dj,MoralesUlion:2015tve,daSilva:2019iwn,Erdas:2023wzy,Cheng:2022mwd}. Therefore, the Casimir effect provides a useful setting in which the interplay between anisotropic scaling and boundary conditions can be explored.

The geometry of the boundaries provides an additional modification of the vacuum spectrum. In particular, surface roughness changes the allowed field modes and can modify the Casimir interaction even when the mean separation between the boundaries is kept fixed. One approach to treating surface roughness is to consider it as a small-amplitude, allowing derivative contributions to be neglected within a perturbative treatment. The Casimir energy under these geometric considerations has already been investigated in $(2+1)$ dimensions for a Ho\v{r}ava-Lifshitz theory, where the boundary deformation was treated perturbatively within a $\zeta$-function framework \cite{Borquez:2023ajx}. This provides a natural starting point for extending the analysis to $(3+1)$ dimensions and to quantum corrections in these models.

For self-interacting scalar field theories, the effective-action formalism provides a systematic framework for determining quantum corrections to the effective potential \cite{Jackiw:1974cv,book1,Kirsten:2007ev,Kirsten:2010zp}.
One of the most notable consequences of self-interacting fields in spacetimes with non-trivial topology is the generation of a topological mass through quantum corrections to the effective potential induced by the underlying topology \cite{Toms:1979ij,Toms:1980sx,Porfirio:2019gdy}. In this framework, quantum loop corrections to the Casimir energy and the topological mass have been extensively studied under various boundary conditions and external fields \cite{Junior:2023feu,Junior:2025thl}. More recently,
quantum corrections have also been investigated in other Lorentz-violating scalar field theories, further highlighting the relevance of Lorentz symmetry breaking in quantum vacuum phenomena \cite{Valuyan:2025vol,Cruz:2020zkc,FariasJunior:2022qsp,Junior:2024smu}. In particular, the Casimir energy and topological mass of a self-interacting scalar field have been analyzed in a Ho\v{r}ava-Lifshitz model with parallel Dirichlet boundaries, including the leading contribution of the quartic self-interaction \cite{Farias:2024uzf}. Following the same approach, in a recent study we analyzed quantum corrections by considering plates with perturbative roughness in the low-temperature regime \cite{Borquez:2026nos}. There, the boundary deformation was treated within a smooth-surface approximation, and the corresponding quantum corrections to the Casimir energy and topological mass were investigated. The present work extends this analysis by incorporating the anisotropic scaling characteristic of a Ho\v{r}ava-Lifshitz theory and by determining the leading contribution of self-interaction to the Casimir energy at two-loop order. This distinction is important because the quartic coupling does not contribute to the Casimir energy at one-loop order in the present formulation, so that the first interaction-dependent contribution arises from the two-loop \emph{figure-eight} vacuum diagram.

In this work, we investigate the Casimir effect of a massive, self-interacting real scalar field in a $(3+1)$-dimensional Ho\v{r}ava-Lifshitz-type theory with rough Dirichlet boundaries at low temperature. We employ the effective-action formalism and represent the functional determinant of the fluctuation operator in terms of a generalized spectral $\zeta$-function. The boundary deformation is treated perturbatively, with the dominant contribution arising from the local modification of the plate separation, while derivative terms associated with lateral variations are suppressed. To extract the finite geometry-dependent contribution to the $\zeta$-function, we employ the Abel-Plana summation formula. In the low-temperature regime, the thermal contribution remains convergent and is exponentially dominated by the lowest eigenvalues, allowing us to retain only the leading terms in the mode sum. Thus, the thermal contribution plays an important role in the quantum corrections in the low-temperature regime. Within this framework, we determine the one-loop Casimir energy and topological mass and analyze the corresponding vacuum stability. Then, we calculate the leading interaction-dependent correction to the Casimir energy at two-loop order through the \emph{figure-eight} vacuum diagram. In addition, we investigate the massless limit and the convergence of the resulting expressions. In this limit, the topological mass and the two-loop Casimir interaction remain finite only for $z=1$, while odd values $z>1$ exhibit infrared divergences associated with the absence of an intrinsic mass scale. Thus, the standard relativistic scaling is recovered in the massless theory, whereas the massive theory retains a characteristic dependence on the anisotropic scaling exponent.

The paper is organized as follows. In Sec.~II, we formulate the Ho\v{r}ava-Lifshitz-type scalar field theory and describe the boundary configuration with spatially dependent roughness. In Sec.~III, we determine the spectrum to first order in the boundary roughness using perturbation theory and construct the corresponding generalized $\zeta$-function. In Sec.~IV, we derive the renormalized effective potential and analyze the resulting Casimir energy, topological mass, and vacuum stability, including the two-loop contribution to the Casimir energy. Finally, Sec.~V presents a summary of our main results and concluding remarks.


\section{Rough plates in Ho\v{r}ava-Lifshitz theory}

We consider rough plates that satisfy Dirichlet boundary conditions and are embedded in a $(3+1)$-dimensional spacetime within a Ho\v{r}ava-Lifshitz-type framework. A crucial feature of the theory is the anisotropic scaling between temporal and spatial coordinates $[t]=-z\,, [x^i]=-1\,,$ where $z$ denotes the anisotropic scaling exponent. This scaling distinguishes temporal and spatial directions and leads to the breaking of Lorentz invariance in the ultraviolet regime. In this framework, spacetime is endowed with a preferred foliation into constant-time spatial hypersurfaces, which is preserved by the residual symmetry group of foliation-preserving diffeomorphisms $\delta t = f(t)$, $\delta x^i=\zeta^i(t,x^k)$, where time is reparameterized independently of the spatial coordinates, while the spatial coordinates may depend on both space and time. The preferred foliation therefore constitutes an essential geometric structure of the theory, for which the Arnowitt-Deser-Misner (ADM) variables provide a natural parameterization. The anisotropic structure of Ho\v{r}ava-Lifshitz theory modifies the set of operators compatible with the reduced symmetry. In particular, the anisotropic scaling allows for higher-order spatial derivatives while keeping the time evolution at second order, providing a natural framework for extending this type of model to relativistic field theories. In the present work, we consider a real scalar field with anisotropic scaling, whose Lagrangian density provides a Ho\v{r}ava-Lifshitz-type generalization of the relativistic Klein-Gordon theory. The corresponding action is given by \footnote{We consider the signature $(+,-,-,-)$}
\begin{eqnarray}
    S
    =
    \int dt\, d^{3}xN\sqrt{g}
    \left(
    \frac{1}{2}\partial_t\Phi\,\partial_t\Phi
    - \frac{1}{2} \xi^{2(z-1)}
    \nabla_{i_1}\cdots\nabla_{i_z}\Phi
    \nabla^{i_1}\cdots\nabla^{i_z}\Phi
    - V\left(\Phi\right)
    \right)
    \,,
\end{eqnarray}
where $V(\Phi)$ is the scalar self-interaction potential, which takes the form
\begin{eqnarray}
    V(\Phi)
    =
    \frac{1}{2}m^2\Phi^2
    +
    \frac{g}{4!}\Phi^4
    \,.
    \label{Potential}
\end{eqnarray}
The corresponding equation of motion is
\begin{eqnarray}
    \left(
    \partial_t^{\,2}
    +
    (-1)^z \xi^{2(z-1)}\nabla_{i_1}
    \cdots\nabla_{i_z}
    \nabla^{i_1}\cdots\nabla^{i_z}
    + m^2
    + \frac{g}{6}\Phi^2
    \right)\Phi
    =0
    \,,
\label{KGeq}
\end{eqnarray}
where $\xi$ is a parameter with dimensions of inverse mass\footnote{In this work, we adopt natural units in which $\hbar = c = 1$.}. The higher-order spatial derivative operator can be written as
\begin{equation}
    \nabla_{i_1}\cdots
    \nabla_{i_z}
    \nabla^{i_1}\cdots
    \nabla^{i_z}
    =
    \Delta^z
    \,,
\label{OLB}
\end{equation}
where $\Delta$ denotes the Laplace-Beltrami operator associated with the spatial metric. Next, we specify the geometry of the irregular boundaries. To that end, we consider a region bounded by two surfaces defined in Cartesian coordinates $(x,y,w)$, where $0\leq w\leq a + f(x,y)$, with $(x,y)\in\mathbb{R}^{2}$. The term $a$ denotes the mean separation between the boundaries and $f(x,y)$ describes a deformation of the upper surface. We assume that the surface deformation is small compared to the mean separation, $|f|\ll a$. To impose the boundary conditions on fixed coordinate surfaces, we introduce the transformation
\begin{equation}
    w =
    \rho\left(
    1
    + \frac{f(x,y)}{a}
    \right)\,,
    \qquad
    0\leq\rho\leq a
    \,.
\end{equation}
This transformation maps the deformed boundary onto the fixed surface $\rho=a$, transferring the effect of the surface roughness to the spatial metric in the transformed coordinates. The resulting metric is\footnote{The perturbations at both plates can be configured such that their contributions are expressed as a linear combination of the respective perturbations; thus, the combined effect of the perturbations at both plates can be equivalently represented by an effective perturbation on a single plate, with its contribution given by the sum of the individual contributions.
}
\begin{eqnarray}
    g_{ij} 
    &=& 
    \left(
    \begin{array}{ccc}
    1 + 
    \left(
    \rho/a
    \right)^2\left(\partial_x f\right)^2
    & \left(
    \rho/a
    \right)^2 \partial_x f \partial_yf 
    & \frac{\rho}{a} \left(1+\frac{f}{a}\right) \partial_xf
    \\
    \left(
    \rho/a
    \right)^2 \partial_x f \partial_y f  
    & 1 + 
    \left(
    \rho/a
    \right)^2
    \left(\partial_yf\right)^2 
    & \frac{\rho}{a} 
    \left(
    1+\frac{f}{a}
    \right) \partial_yf
    \\
    \frac{\rho}{a}
    \left(
    1+\frac{f}{a}
    \right) \partial_xf
    & \frac{\rho}{a}  
    \left(
    1+\frac{f}{a}
    \right) \partial_yf
    & \left(
    1+\frac{f}{a}
    \right)^2 
    \\
    \end{array}
    \right)
    \,.
    \label{metric}
\end{eqnarray}
The Laplace-Beltrami operator associated with the metric of Eq.~(\ref{metric}), and acting on the scalar field, is given by
\begin{eqnarray}
    \Delta\Phi
    &=&
    \Delta_x\Phi
    + \Delta_y\Phi
    + \frac{1}{(a+f)^2}\left[
    \rho^{2}\left(\left(\partial_{x}f\right)^{2} + \left(\partial_{y}f\right)^{2}\right)
    + a^2\right]\Delta_{\rho}\Phi
    \nonumber
    \\&&
    + \frac{\rho}{\left(a+f\right)^2}\left[
    2\left(\left(\partial_{x}f\right)^{2} + \left(\partial_{y}f\right)^{2}\right)
    - \left(a+f\right)\left(\Delta_{x}f + \Delta_{y}f\right)
    \right]\partial_{\rho}\Phi
    \nonumber
    \\&&
    - \frac{2\rho}{a+f}\left(
    \partial_{x}f\Delta_{x\rho}\Phi
    + \partial_{y}f\Delta_{y\rho}\Phi\right)
    \,.
    \label{L-B-O}
\end{eqnarray}
It is convenient to introduce dimensionless coordinates according to the following
\begin{eqnarray}
    \begin{split}
    x & = u_1 L_1\,, \qquad -1/2\leq u_1\leq 1/2\,,
    \\
    y & = u_2 L_2\,, \qquad  -1/2\leq u_2\leq 1/2\,,
    \\
    \rho & = va \,,\qquad\quad  \qquad 0\leq v\leq 1\,,
    \end{split}
    \label{finalchangeofvariable}
\end{eqnarray}
where $L_1$ and $L_2$ characterize their lateral extensions. In the new coordinates, we denote the surface profile by
\begin{equation}
    f(u_1L_1,u_2L_2)
    \equiv
    \hat f(u_1,u_2)
    \,.
\end{equation}
(From now on, for the components $u_{1}$ and $u_{2}$, we will work with a compact formulation represented by $u_{i}$ with $i=1,2$, so $\hat{f}(u_{i})=f(u_{i}L_{i})$. Similarly, the spatial operator of these coordinates will be defined by $\partial^{2}_{u_{1}} + \partial^{2}_{u_{2}}=\partial^{2}_{u_{i}}=\Delta_{u_{i}}$). We consider weakly deformed plates characterized by the small-amplitude conditions $a/L_i\ll1$,
$
|\hat f|/a\ll1
$ and $
|\partial\hat f|/a\ll1
$. 
These conditions ensure that the surface deformation is small in amplitude, allowing for a perturbative treatment, and derivative contributions
such as $\partial_{i}\hat{f}$ and $\partial_{i}^{2}\hat{f}$ are suppressed with respect to the leading geometric corrections associated with the local deformation amplitude. Since the plates are assumed to be laterally extended compared with their separation, we set
$
L_1=L_2=L,
$
and neglect edge effects by taking the limit $L\rightarrow\infty$. Under these assumptions, the operator in Eq.~\eqref{L-B-O} can be decomposed into a free part and perturbative contributions, which are treated as effective perturbation potentials
\begin{eqnarray}
    \Delta\Phi
    =
    \left(\frac{1}{L^2}\Delta_{u_i}
    + \frac{1}{a^2}\Delta_v
    - \mathcal{M}(u_i)\Delta_v\right)\Phi
    \,,
\label{Delta}
\end{eqnarray}
where the expansion of $\mathcal{M}$ is defined by
\begin{eqnarray}
    \mathcal{M}(u_i) = 
    2\frac{\hat{f}(u_{i})}{a^{3}}
    - 3\frac{\hat{f}^{2}(u_{i})}{a^4}
    + \mathcal{O}(\hat{f}^{3})
    \,.
    \label{M}
\end{eqnarray}
The operator \eqref{Delta}, together with Eqs. \eqref{M} and \eqref{Potential}, will be used below to construct the spectrum of scalar-field fluctuations in Ho\v{r}ava-Lifshitz theory.


\section{Generalized $\zeta$-function}

\subsection{Effective action}

To determine the one- and two-loop corrections
generated by the one-particle-irreducible diagrams, we
construct the effective action within the loop-expansion framework \cite{book1}. The effective action $\Gamma[\Psi]$ incorporates both the classical dynamics and quantum corrections arising from fluctuations about the background field $\Psi$, which is defined as the expectation value of the quantum field in the presence of external sources. Its loop expansion is given by
\begin{equation}
    \Gamma[\Psi]
    = 
    S[\Psi]
    + \sum_{i=1}\hbar^i\Gamma^{(i)}[\Psi]
    \,,
\end{equation}
where $S[\Psi]$ is the classical action and $\Gamma^{(i)}[\Psi]$ represents the $i$-loop contribution. The background-field decomposition is introduced as $\Phi =\Psi + \varphi$, where $\varphi$ denotes the quantum fluctuations around the background configuration. Expanding the action to second order in the quantum field yields
\begin{equation}
    S\left[\Psi+\varphi\right]
    = 
    S\left[\Psi\right]
    +
    \frac{1}{2}
    \varphi S_{2}\left[\Psi\right] \varphi
    +
    \mathcal{O}\left(\varphi^3\right)
    \,,
  \end{equation}
where $S_{2}[\Psi]$ corresponds to the second functional derivative of the classical action evaluated in the background field. Thus, at one-loop order, the quantum corrections are fully determined by the quadratic fluctuation operator $S_{2}[\Psi]$.
Performing the Gaussian functional integral,
\begin{eqnarray}
    \exp\left( i \Gamma^{(1)}\left[\Psi\right] \right)
    =
    \int \mathcal{D}\varphi
    \exp\left( \frac{i}{2}\int \,dt\,d^3x\,
    \varphi\,S_2\left[\Psi\right]\varphi \right)
    \,,
\end{eqnarray}
yields the standard one-loop result
\begin{eqnarray}
    \Gamma^{(1)}\left[\Psi\right]
    = 
    \frac{i}{2}\ln \det S_{2}\left[\Psi\right]
    \,.
\end{eqnarray}
To extract physical information from the effective action,
such as the vacuum structure, we consider the effective
potential, $V_{\text{eff}}$ , defined as the effective action evaluated for constant background fields. In this limit, all derivatives acting on the background vanish, and the effective action reduces to
\begin{eqnarray}
    \Gamma[\Psi]
    = 
    \int d\mathcal{V}_{E}
    \left[
    - V_{\mathrm{eff}}\left(\Psi\right)
    + \mathcal{O}\left(\partial \Psi\right)
    \right]
    \,.
\end{eqnarray}
The leading term in the derivative expansion is determined entirely by the effective potential, which admits a loop expansion of the form $V_{\mathrm{eff}}(\Psi)
=
\sum_{i=0}V^{(i)}_E\left(\Psi\right)$, 
where $V^{(0)}_E$ is the classical potential, and the one-loop Euclidean effective potential is given by\footnote{We perform the Wick rotation, $t=-i\tau$, thereby formulating the functional integral in Euclidean spacetime.}
\begin{equation}
    V^{(1)}_E\left(\Psi\right)
    = -\frac{1}{\mathcal{V}_E}
    \ln\int\mathcal{D}\varphi
    \exp\left(
    - \frac{1}{2}\int d\tau d^3x\,\varphi\,S_2\left[\Psi\right]\varphi\right)
    \,,
\end{equation}
where $\mathcal{V}_E=\int d\tau d^{3}x\sqrt{g}$ is the Euclidean volume and $S_{2}[\Psi]$ represents
the Euclidean elliptic operator.

For the problem we present here, we must construct this elliptic operator from \eqref{KGeq}, which we will call $\hat{Q}$. Therefore, considering \eqref{OLB}, the total operator is given by
\begin{equation}
\label{S2P}
    \hat{Q}
    = 
    - \partial^2_\tau
    + \mathcal{\hat{P}}
    \,,\qquad 
    \mathcal{\hat{P}}
    =
    (-1)^{z} l^{2(z-1)}
    \Delta^z
    +
    M^2
    \,.
\end{equation}
where 
\begin{eqnarray}
    M^{2}
    =
    m^{2}+\frac{g}{2}\Psi^{2}\,.
\end{eqnarray}
The operator $\hat{Q}$ encodes the characteristic structure of the anisotropic spatial-derivative through the scaling exponent. As a result, quantum fluctuations are governed by the corresponding modified dispersion relation, which depends on the background field $\Psi$. In our case, the explicit form of the volume is determined using the spatial metric \eqref{metric}, that is,
\begin{eqnarray}
\label{volume}
    \mathcal{V}_E
    =
    \beta L^2\left(
    a 
    + \int_{-1/2}^{1/2}
    \int_{-1/2}^{1/2}\hat{f}(u_i)\,du_i
    \right)\,.
\end{eqnarray}

The quantum contribution is obtained from the functional determinant of the Euclidean operator subject to the boundary conditions imposed by the parallel plates. We regularize this determinant by introducing the generalized $\zeta$-function,
\begin{equation}
    \ln \det \hat{Q}
    = 
    - \left.\zeta'_{\hat{Q}}(s) \right|_{s=0}
    \,,
    \qquad
    \zeta_{\hat{Q}}(s)
    = \sum_{n} \Omega_{n}^{-s}
    \,,
\end{equation}
where the set $\{\Omega_n\}$ are the eigenvalues of the elliptic operator. Thus, the one-loop effective potential is given by
\begin{eqnarray}
    V^{(1)}_E(\Psi)
    =
    - \frac{1}{2\mathcal{V}_E}
    \left[
    \zeta'(0)
    + \zeta(0)\ln\mu^2
    \right],
\label{E-V-ZF}
\end{eqnarray}
where $\mu$ is an arbitrary mass scale introduced to render the generalized $\zeta$-function dimensionless. At this stage, the quantum problem is encoded in the generalized $\zeta$-function, and its spectrum incorporates the effects of anisotropic scaling, plate geometry, and finite temperature. We will discuss the two-loop order correction to the theory and its functional dependence later.

The resulting effective potential generally contains divergent contributions, which must be removed through an appropriate renormalization procedure. We impose the renormalization conditions
\begin{eqnarray}
\label{Rconditions}
    \left.\frac{d^4 V_{\mathrm{eff}}(\Psi)}
    {d\Psi^4}\right|_{\Psi=\hat{\Psi}} 
    = g,
    \qquad
    \left.\frac{d^2 V_{\mathrm{eff}}(\Psi)}
    {d\Psi^2}\right|_{\Psi=\Psi_0} 
    = m^2,
\end{eqnarray}
where $\hat{\Psi}$ is a mass-dimensional field and $\Psi_0$ denotes the vacuum configuration, determined by $
V_{\mathrm{eff}}'(\Psi_0)=0$, together with the stability condition $V_{\mathrm{eff}}''(\Psi_0)>0.$ The latter conditions ensure that $\Psi_0$ is a stable local minimum and therefore defines the vacuum state. Additionally, we fix the normalization of the vacuum energy by requiring $V_{\mathrm{eff}}(\Psi_0)=0$. Once the effective potential has been renormalized and the vacuum configuration is identified, the quantum contribution to the Casimir energy and topological mass can be extracted from the resulting effective potential. 

In the following subsection, we analyze the spectrum of the operator by means of perturbation theory and its anisotropic, geometry, and temperature dependence.


\subsection{The perturbation theory}

To determine the spectrum of the spatial operator, we employ perturbation theory to first order in the eigenvalue corrections. The corresponding eigenvalue problem must be supplemented by appropriate boundary conditions; in our case, we consider Dirichlet boundary conditions. Thus, the anisotropic scalar field is subject to
\begin{eqnarray}
    \Phi\left(\tau,u_{i},0\right)
    =
    \Phi\left(\tau,u_{i},1\right)
    =
    0\,.
\label{DBC}
\end{eqnarray}
For $\tau\in\mathbb{C}$, the eigenfunctions of the Euclidean  operator \eqref{S2P} can be written as
\begin{eqnarray}
    \Phi_{l,k_{i},n}(\tau,u_{i},v)
    =
    \frac{1}{\beta}e^{\frac{2\pi i l}{\beta}\tau}\phi_{k_{i},n}(u_{i},v)\,,\quad l\in\mathbb{Z}\,.
\end{eqnarray}
The total eigenvalues of $\hat{Q}$ take the form
\begin{eqnarray}
    \Omega_{l,k_{i},n}
    = 
    \left( \frac{2\pi l}{\beta} \right)^{2}
    + \lambda_{k_{i},n}^{2}\,,
\end{eqnarray}
where $\lambda_{k_{i},n}$ denotes the spectrum of the spatial operator $\hat{\mathcal{P}}$ (see Eq. (\ref{S2P}))\footnote{As was done previously for the coordinates, we compact the eigenvalues $k_{1},k_{2}$ associated with these coordinates in the form $k_{i}$, with $i=1,2$, therefore $k_{i}^{2}=k_{1}^{2} + k_{2}^{2}$.}. The corresponding generalized $\zeta$-function is
\begin{eqnarray}
    \zeta_{\hat{Q}}\left(s\right)
    =
    \sum_{l= -\infty}^{\infty} \sum_{k_{i}, n=1}^{\infty}
    \left[
    \left( \frac{2\pi l}{\beta} \right)^{2}
    + \lambda_{k_{i},n}^{2}
    \right]^{-s}
    \,.
\end{eqnarray}
To account for the modification of the spatial operator induced by the roughness of the plates, the spatial spectrum is determined perturbatively. We first consider the unperturbed configuration of parallel plates and subsequently incorporate the surface profile through the corresponding perturbation operator. The unperturbed problem is given by
\begin{eqnarray}
    -\Delta\phi^{(0)}
    =
    -\left(
    \frac{1}{L^2}\Delta_{u_i}
    + \frac{1}{a^2}\Delta_{v}
    \right)\phi^{(0)}=\lambda^{(0)}\phi^{(0)}\,,
\label{Delta_0}
\end{eqnarray}
whose normalized eigenfunctions are given by
\begin{eqnarray}
    \phi^{(0)}_{k_{i},n}(u_{i},v)
    =
    \sqrt{2}\sin\left(n\pi v\right)
    e^{ik_{i}u_{i}}
    \,,
\label{sol_phi_0}
\end{eqnarray}
where $k_{i}=\frac{\pi q_{i}}{L}$ with $q_{i}\in\mathbb{Z}$, and $n\in\mathbb{N}$. The corresponding eigenvalues are
\begin{eqnarray}
    \lambda_{k_i,n}^{(0)}
    =
    \left(\frac{n\pi}{a}\right)^2
    +
    k_{i}^2
    \,.
\end{eqnarray}
The first order correction is obtained in terms of the perturbation operator (\ref{M}), it is given by\footnote{ One of the aims of this work is to determine the contribution of rough plates to quantum corrections. To this end, we will focus on perturbations in $\hat{f}/a$ up to at least the second order. However, in perturbation theory, there is no guarantee that contributions at the second order of $\hat{f}/a$ do not arise at the second order of the $\lambda$ perturbation. Therefore, we will proceed with this analysis assuming, as noted in \cite{Borquez:2023ajx}, that the function $\hat{f}$ has periodic behavior, thereby ensuring that no further terms arising from $\lambda^{(2)}$ need to be considered in the perturbation expansion.}
\begin{eqnarray}
    \lambda_{k_i,n}^{(1)}
    =
    \int_{-1/2}^{1/2}
    \int_{-1/2}^{1/2}
    \int_0^1 du_i\,dv\,
    \Phi^{(0)*}_{k_{i}, n}(u_{i},v)
    \mathcal{M}(u_i)\,
    \Delta_{v}\Phi_{k_{i}, n}^{(0)}(u_{i},v)\,.
\end{eqnarray}
Retaining terms up to first order in the perturbation theory, the total spectrum is given by
\begin{eqnarray}
\label{TotalEV}
    \lambda_{k_i,n}
    =
    \left(\frac{n\pi}{a}\right)^2
    + k_{i}^2
    -\left(n\pi\right)^2
    \int_{-1/2}^{1/2}
    \int_{-1/2}^{1/2}\,du_i\,\mathcal{M}\left(u_i\right)
    \,.
\end{eqnarray}    
Substituting the perturbative spectrum into the generalized $\zeta$-function and taking the limit $L\rightarrow\infty$, the sums over the periodic directions become continuous integrals $\sum_{k_{i}}\rightarrow\left(\frac{L}{2\pi}\right)\int dk_{i}$. Using the integral representation of the Gamma function, the generalized $\zeta$-function can be written as 
\begin{eqnarray}
\label{ZetaP_continuo_Generalized}
    \zeta_{\hat{Q}}(s)
    &=&
    \frac{1}{\Gamma\left(s\right)}\left(\frac{L}{2\pi}\right)^2
    \sum_{n,l=1}^{\infty}
    \int_{-\infty}^{\infty}
    \int_{-\infty}^{\infty}
    \,dk_i
    \nonumber
    \\
    &&\times
    \int_0^{\infty}dt\,t^{s-1}
    \exp\left\{
    - t\left[
    \left(
    \frac{2\pi l}{\beta}\right)^2
    + \xi^{2(z-1)}\left(
    \mathcal{R} n^2
    + k_{i}^{2}
    \right)^{z}
    + M^2 \right]
    \right\},
\end{eqnarray}
where all the information regarding the roughness of the plates is condensed into the following expression
\begin{eqnarray}\label{R12}
    \mathcal{R}
    =
    \pi^2\left(
    \frac{1}{a^2}
    - \int_{-1/2}^{1/2}
    \int_{-1/2}^{1/2}\,du_i\,\mathcal{M}(u_i)
    \right)
    \,.
\end{eqnarray}
Starting from the generalized $\zeta$-function \eqref{ZetaP_continuo_Generalized}, we evaluate the integrations over the continuous momenta parallel to the plates by introducing polar coordinates in the $k_i$ plane, that is, $k_{1}^{2} + k_{2}^{2}\rightarrow k^{2}$. A suitable change of variables reduces the resulting radial integral to the standard form of the incomplete Gamma function (see \eqref{IGF}). Thus,
\begin{eqnarray}
    \iint dk_i\,
    e^{-t\,\xi^{2(z-1)} 
    \left(\mathcal{R} n^2 + k_i^{2} \right)^z }
    =
    \frac{\pi\,t^{-1/z}}{z\,\eta_{z}}
    \Gamma\left(
    z^{-1}, t\,(\eta_{z}\,\mathcal{R} n^2)^z \right)
    \,,
\end{eqnarray}
where $\eta_{z}=\xi^{2(1 - \frac{1}{z})}$. To further analyze the resulting expression, we apply the Poisson resummation formula to the sum over $l$, which yields
\begin{eqnarray}\label{ZetaP_continuo_Generalized_withoutk_12_PoisonR}
    \zeta_{\hat{Q}}(s)
    &=&
    \frac{\beta\sqrt{\pi}}{z\eta_{z}\Gamma(s)}
    \left(\frac{L}{2\pi}\right)^2
    \sum_{n=1}^{\infty}
    \int_0^{\infty}dt\,
    t^{s-\frac{3}{2}-\frac{1}{z}}
    \nonumber\\
    &&
    \times\,
    \Gamma\left(z^{-1}, t\,(\eta_{z}\,\mathcal{R} n^2)^z\right)
    e^{-tM^{2}}
    \left(
    \frac{1}{2}
    +
    \sum_{l=1}^{\infty}
    e^{-\frac{\beta^2l^2}{4t}}
    \right).
\end{eqnarray}
The generalized $\zeta$-function can be decomposed into two contributions: the first is associated solely with the geometry of the plates, while the second relates to both geometry and temperature. For the purely geometric contribution, the incomplete Gamma function can be expressed in terms of the confluent hypergeometric function by using the identity
\begin{eqnarray}
    \Gamma\left(\alpha, x\right) 
    = 
    e^{-x}\,
    \Psi\left(1-\alpha,\,1-\alpha;\,x\right)
    \,,
\end{eqnarray}
where
\begin{eqnarray}
    \Psi(\alpha,\gamma;z)
    =
    \frac{\Gamma(1-\gamma)}{\Gamma(\alpha-\gamma+1)}\,
    {}_1F_1\left(\alpha;\gamma;z\right)
    +
    \frac{\Gamma(\gamma-1)}{\Gamma(\alpha)}\,
    z^{1-\gamma}\,
    {}_1F_1\left(\alpha-\gamma+1;\,2-\gamma;\,z\right)
    \,,
    \nonumber\\
\end{eqnarray}
and ${}_1F_1$ denotes the confluent hypergeometric function (see \cite{GradshteynRyzhik}).
We first evaluate the integral over the variable $t$. The first contribution is obtained using the Laplace transform
\begin{eqnarray}
    \int_{0}^{\infty} t^{b-1}\,
    \Psi(a,c;t)\,
    e^{-\varrho t}\,dt
    =
\frac{\Gamma(b)\,\Gamma(b-c+1)}
    {\Gamma(a+b-c+1)}\;
    \varrho^{-b}\;
    {}_2F_{1}\!\left(a,\,b;\,a+b-c+1;\,1-\varrho^{-1}\right)
    \,,
    \nonumber\\
\end{eqnarray}
valid for $\Re(\varrho)>1/2$. The second contribution in \eqref{ZetaP_continuo_Generalized_withoutk_12_PoisonR} can be analyzed using the integral representation of the incomplete Gamma function,
\begin{eqnarray}
    \Gamma\left(\alpha,\gamma\right)
    =
    \int_{\gamma}^\infty 
    \vartheta^{\alpha - 1}e^{-\vartheta}\,d\vartheta
    \,,
    \label{IGF}
\end{eqnarray}
by considering $\vartheta=\omega\,t\left(\eta_{z}\mathcal{R}n^{2}\right)^{z}$, which yields
\begin{eqnarray}
\label{EcsumT}
    \sum_{n,l=1}^{\infty}
    \eta_{z}\mathcal{R}n^{2}
    \int_1^\infty
    d\omega \,\omega^{\frac{1}{z}-1}
    \int_0^\infty
    dt\,t^{s-\frac{3}{2}}
    \exp\left[-t\left(\omega(\eta_{z}\mathcal{R}n^{2})^{z} 
    + M^2 \right)
    - \frac{\beta^2l^{2}}{4t}\right]
    \,.
\end{eqnarray}
The resulting integral with respect to $t$, given a suitable change of variables, takes the standard form associated with the modified Bessel function of the second kind. In the low-temperature regime, $\beta\gg a$, the argument of the Bessel function becomes large, allowing us to use the asymptotic expansion
\begin{eqnarray}
    K_\nu(x)
    \sim
    \sqrt{\frac{\pi}{2x}}e^{-x},
    \qquad x\gg1.
\end{eqnarray}
Therefore, the higher-order terms are exponentially suppressed, and the thermal contribution is dominated by the lowest eigenvalues. Consequently, the leading temperature-dependent contribution is obtained by retaining only the lowest modes, $n=l=1$, while contributions with $n,l>1$ can be neglected in the order considered here. Then, Eq.~\eqref{EcsumT} takes the form
\begin{eqnarray}
\label{zetaTlimit}
    \eta_{z}\mathcal{R}\sqrt{\pi}
    \left(\frac{\beta}{2}\right)^{s-1}
    \int_1^\infty d\omega\,
    \omega^{\frac{1}{z}-1}
    \left(\omega(\eta_{z}\mathcal{R})^{z} + M^{2}\right)^{-\frac{s}{2}}
    e^{-\beta\sqrt{\omega(\eta_{z}\mathcal{R})^{z} + M^{2}}}
    \,.
\end{eqnarray}
Finally, the generalized $\zeta$-function relevant to our analysis is given by
\begin{eqnarray}
\label{Zetafunctionwork}
    \zeta_{\hat{Q}}(s)
    &=&
    \frac{\beta\sqrt{\pi}}{z\eta_{z}\Gamma(s)}
    \left(
    \frac{L}{2\pi}
    \right)^2
    \left[
    \frac{
    \Gamma\left(
    s - \frac{1}{2}
    \right)}
    {2\left(
    s - \frac{1}{2} - \frac{1}{z}
    \right)}
    \sum_{n=1}^{\infty}
    \left((\eta_{z}\mathcal{R})^{z}n^{2z}
    + M^{2}\right)^{-s+\frac{1}{2}+\frac{1}{z}}
    \right.
    \nonumber\\
    &&
    \times
    {}_2F_{1}\left(
    1 - \frac{1}{z},
    s - \frac{1}{2} - \frac{1}{z};
    s + \frac{1}{2} - \frac{1}{z};
    \frac{M^{2}}{(\eta_{z}\mathcal{R})^{z}n^{2z} + M^{2}}\right)
    \nonumber\\
    &&
    \left.
   + \eta_{z}\mathcal{R}\sqrt{\pi}
    \left(\frac{\beta}{2}\right)^{s-1}
    \int_1^\infty d\omega\,
    \omega^{\frac{1}{z}-1}
    \left(\omega(\eta_{z}\mathcal{R})^{z} + M^{2}\right)^{-\frac{s}{2}}
    e^{-\beta\sqrt{\omega(\eta_{z}\mathcal{R})^{z} + M^{2}}}
    \right]\,.
    \nonumber\\
\end{eqnarray}
The resulting $\zeta$-function contains a divergent sum over the quantum number $n$ in its first contribution, therefore, we employ the Abel-Plana regularization method to extract its finite part. We emphasize that this procedure is applied only to the purely geometric sector, since the temperature-dependent sector has already been simplified by adopting the low-temperature approximation. The Abel-Plana decomposition separates the result into three contributions: a bulk term associated with the vacuum, a boundary term, and a finite term arising from the boundary conditions imposed by the plates, in the following way
\begin{eqnarray}
    \sum_{n=0}^{\infty} f(n)
    &=&
    \int_{0}^{\infty} f(x)\,dx
    + \frac{f(0)}{2}
    + i\int_{0}^{\infty} \frac{f(ix)-f(-ix)}{e^{2\pi x}-1}\,dx
    \,.
\end{eqnarray}
The latter contains the physically relevant contribution and is the term used in the calculation of the effective potential. Therefore, the finite part of the generalized $\zeta$-function is given by
\begin{eqnarray}
\label{Zetafunctionwork-Abel-Plana}
    \zeta_{\hat{Q}}(s)
    &=&
    \frac{\beta\sqrt{\pi}}{z\eta_{z}\Gamma(s)}
    \left(
    \frac{L}{2\pi}
    \right)^2
    \left\{
    -\frac{
    \Gamma\left(
    s - \frac{1}{2}
    \right)}
    {\left(
    s - \frac{1}{2} - \frac{1}{z}
    \right)}
    \right.
    \nonumber\\
    &&
    \times
    \sin\left[
    \pi\left(-zs+\frac{z}{2}+1\right)\right]
    \int_{M^{\frac{1}{z}}(\eta_{z}\mathcal{R})^{-\frac{1}{2}}}^\infty
    \frac{\left((\eta_{z}\mathcal{R})^{z}x^{2z}
    - M^2\right)^{-s+\frac{1}{2}+\frac{1}{z}}}
    {e^{2\pi x}-1}
    \nonumber\\
    &&
    \times
    {}_2F_{1}\left(
    1 - \frac{1}{z},
    s - \frac{1}{2} - \frac{1}{z};
    s + \frac{1}{2} - \frac{1}{z};
    \frac{M^2}{-(\eta_{z}\mathcal{R})^{z}x^{2z}+M^2}\right)
    \,dx
    \nonumber\\
    &&
    \left.
    + \eta_{z}\mathcal{R}\sqrt{\pi}
    \left(\frac{\beta}{2}\right)^{s-1}
    \int_1^\infty d\omega\,
    \omega^{\frac{1}{z}-1}
    \left(\omega(\eta_{z}\mathcal{R})^{z} + M^{2}\right)^{-\frac{s}{2}}
    e^{-\beta\sqrt{\omega(\eta_{z}\mathcal{R})^{z} + M^{2}}}
    \right\}\,.
    \nonumber\\
    \label{ZetaAP}
\end{eqnarray}
For the purely geometrical sector, the finite Abel-Plana contribution arises from the discontinuity between the two analytic continuations across the relevant branch cut. For $0\leq x< M^{\frac{1}{z}}(\eta_z\mathcal{R})^{-\frac{1}{2}}$, the argument of the fractional power in the integral remains real and positive, and therefore no phase difference arises between the continuations. In contrast, for $x\geq M^{\frac{1}{z}}(\eta_z\mathcal{R})^{-\frac{1}{2}}$, the argument reaches the branch cut, and the continuations approaching it from opposite sides acquire different phases, giving rise to a finite contribution. For integer $z$, this contribution vanishes for even $z$, whereas it remains non-vanishing for odd $z$. This is reflected in the sine function, which depends on $s$ and $z$. In our case, we will eventually consider $s=0$.


\section{ Renormalized Effective Potential, Casimir Energy, and Topological Mass}\label{SectionE-P}

\subsection{One-loop correction}

Having obtained the generalized $\zeta$-function \eqref{ZetaAP}, we now proceed to compute the one-loop effective potential defined in Eq.~\eqref{E-V-ZF}. Expanding the generalized $\zeta$-function around $s=0$, we find that $\zeta(0)=0$, independently of the anisotropic scaling exponent, while its derivative at $s=0$ is non-vanishing and determines the finite contribution. Thus, the one-loop contribution is entirely determined by $\zeta'(0)$. Therefore, the one-loop effective potential is 
\begin{eqnarray}
    V_{\text{eff}}
    &=&
    \frac{m^{2}}{2}\Psi^{2} 
    + \frac{g}{4!}\Psi^{4} 
    \nonumber\\
    &&
    + \frac{\beta L^{2}}{\mathcal{V}_{E}}
    \left[
    \frac{\sin \left(\frac{\pi}{2} (z+2)\right)}{2\pi\eta_{z}(z+2)}
    \int_{M^{\frac{1}{z}}(\eta_{z}\mathcal{R})^{-\frac{1}{2}}}^\infty
    \frac{\left((\eta_{z}\mathcal{R})^{z}x^{2z}
    - M^2\right)^{\frac{1}{z}+\frac{1}{2}} }
    {e^{2\pi x}-1}
    \right.
    \nonumber\\
    &&
    \times\, _2F_1\left(-\frac{1}{2}-\frac{1}{z},
    1-\frac{1}{z};
    \frac{1}{2}-\frac{1}{z};
    \frac{M^2}{-(\eta_{z}\mathcal{R})^{z}x^{2z}+M^2}\right)
    \,dx
    \nonumber\\
    &&
    \left.
    - \frac{\mathcal{R}}
    {4\pi z \beta}
    \int_1^{\infty } \omega ^{\frac{1}{z}-1}
    e^{-\beta  \sqrt{\omega(\eta_{z}\mathcal{R})^{z} + M^2}} \, d\omega
    \right]
    \,.
    \label{Veff2}
\end{eqnarray}
The effective potential at one-loop order must, in principle, be renormalized. To determine the corresponding counterterms, we implement the renormalization process using the conditions defined in \eqref{Rconditions} and the $a\rightarrow\infty$ limit, indicating that the plates are infinitely separated. For this analysis, we consider the stable vacuum solution to be $\Psi=0$ (this solution will be analyzed later). After applying the conditions \eqref{Rconditions}, we observe that the remaining integrals completely depend on the parameter $a$; thus, the limit $a\rightarrow\infty$ requires careful analysis. Notably, and independently of the outcome of the integrations, we see that in the first integration of \eqref{Veff2} (after performing a change of variables that eliminates the dependence of the integration limits on $a$ and the field), the exponential factor in the integrand suppresses the entire integral in that limit, whereas the temperature factor, the Euclidean volume dividing it, contributes involving $a$ that completely cancels it out. This occurs under all conditions of the normalization process; therefore, counterterms are not needed in this theory. These conclusions remain valid without specifying the explicit form of the function $\hat{f}$. The Abel-Plana method allows for an analytical representation of the purely geometric eigenvalues, which typically encode the ultraviolet structure of the theory. Consequently, the finiteness of the effective potential is achieved through the construction of the spectral $\zeta$-function, together with the renormalization conditions adopted here. This result is also consistent with the anisotropic structure of Ho\v{r}ava-Lifshitz-type theories, which, as has already been seen in the literature \cite{Farias:2024uzf}, prevents the appearance of logarithmic and scale-dependent terms. In the present formulation, this is reflected in the generalized $\zeta$-function, whose contribution vanishes when evaluated at $s=0$ for the considered background. Finally, the result obtained in \eqref{Veff2} is entirely finite and corresponds precisely to the renormalized effective potential $V^{R}_{\text{eff}}$.

The next step is to analyze the vacuum stability of our solution \eqref{Veff2}, which, as mentioned previously, corresponds to the renormalized one-loop effective potential. To this end, we introduce the change of variables
$
x=\theta M^{\frac{1}{z}}(\eta_z \mathcal{R})^{-\frac{1}{2}},
$
in $V_{\text{eff}}$, this removes the explicit mass dependence from the integration limits and allows us to perform a systematic perturbative expansion up to first order in the self-coupling $g$, which we assume to satisfy $g\ll1$. This perturbative treatment is adopted solely for the purpose of analyzing the stability of the system. The stationary condition,
$
V_{\text{eff}}'(\Psi)=0,
$
determines the critical points, which correspond to the possible vacuum states. A critical point represents a stable vacuum if
$
V_{\mathrm{eff}}''(\Psi)>0,
$
ensuring that the effective potential has a local minimum at that point. Accordingly, the equation determining the critical points is given by
\begin{eqnarray}
    0
    &=&
    \Psi\left\{
    m^2
    + \frac{g}{6}\Psi^2 
    + \frac{g\beta L^{2}}{\mathcal{V}_{E}}
    \left[
    \frac{ m^{\frac{3}{z}-1} \mathcal{R}^{-\frac{3}{2}}}{4\pi z\eta^{\frac{5}{2}}_{z}(z+2)}
    \sin \left(\frac{\pi}{2} (z+2)\right)
    \right.
    \right.
    \nonumber\\
    &&
    \times
    \sqrt{\eta_z  \mathcal{R}}
    \int_{1}^\infty
    \left(\theta^{2z}
    - 1\right)^{\frac{1}{z}+\frac{1}{2}}
    \left((z+3)\sqrt{\eta_z  \mathcal{R}}\,\mathcal{I}_{\theta}^{-1}
    \right.
    \nonumber\\
    &&
    \left.
    - 2 \pi  \theta  m^{\frac{1}{z}} 
    \mathcal{I}_{\theta}^{-1}(1 + \mathcal{I}_{\theta}^{-1})\right)
    \, _2F_1\left(-\frac{1}{2}-\frac{1}{z},
    1-\frac{1}{z};
    \frac{1}{2}-\frac{1}{z};
    \frac{1}{-\theta^{2z}+1}\right)
    \,d\theta
    \nonumber\\
    &&
    \left.
    \left.
    +
    \frac{\mathcal{R}}
    {8\pi z }
    \int_1^{\infty } \omega ^{\frac{1}{z}-1}
    \frac{   e^{-\beta  \sqrt{m^2 
    + \omega  (\eta_z \mathcal{R})^z}}}
    {\sqrt{m^2 + \omega (\eta_z \mathcal{R})^z}}\, d\omega
    \right]
    \right\}
    \,.
\label{vacuum-solutions}
\end{eqnarray}
where
\begin{eqnarray}
    \mathcal{I}_{\theta}
    =
    e^{\frac{2\pi \theta m^{1/z}}{\sqrt{\eta_{z}\mathcal{R}}}}
    - 1\,.
\end{eqnarray}
Thus, Eq.~\eqref{vacuum-solutions} admits the trivial solution $\Psi=0$, which corresponds to a stable vacuum, and, in the perturbative regime considered, this solution satisfies $V_{\text{eff}}''(0)>0$. 

Other solutions follow from the quadratic equation in Eq.~\eqref{vacuum-solutions} and correspond to a possible nontrivial vacuum solution. For even values of the anisotropic exponent $z$, the purely geometric contribution vanishes, while the thermal contribution remains finite and strictly positive for all $z>0$. Then, a nontrivial solution can exist only for negative self-coupling and suitable values of the mass and thermal contributions. However, the second derivative of the effective potential evaluated in this solution is negative, which shows that these configurations do not correspond to stable vacuum states. In the massless case, within the low-temperature regime, no real nontrivial solution exists. 

For odd values of $z$, the purely geometric contribution is nonvanishing and does not admit a closed-form expression, unless some of the parameters of the theory are fixed. Therefore, if this is the case, it can modify the vacuum equation and the corresponding critical points, potentially allowing for stable, non-trivial vacuum configurations. In the massless limit, this quadratic solution imposes a constraint on the anisotropic scaling exponent $z$ within the integral of the purely geometric sector. Upon identifying the terms containing the mass in this integral and taking the zero-mass limit, we note the following
\begin{eqnarray}
    \lim_{m\rightarrow 0}
    \left\{m^{\frac{3}{z}-1}\left(
    (z + 3)\sqrt{\eta_z  \mathcal{R}}\,\mathcal{I}^{-1}_{\theta}
    - 2\pi\theta m^{\frac{1}{z}}\mathcal{I}_{\theta}^{-1}
    (1 + \mathcal{I}_{\theta}^{-1}) \right)
    \right\}
    =0
    \,,
\end{eqnarray}
only for  $z<2$.
For odd integer values $z>1$, the purely geometric contribution develops a power-like infrared divergence in the massless limit. This divergence is not induced by the surface roughness, but instead arises from the absence of an intrinsic mass scale that regulates the infrared behavior of the theory. As the anisotropic scaling exponent $z$ increases, the massless theory becomes increasingly sensitive to infrared modes. Thus, the purely geometric contribution remains finite in the massless limit only for $z=1$. In contrast, for massive theory, the presence of the mass scale regularizes the infrared behavior, and it is possible, for suitable choices of the theory parameters, to obtain a real solution. However, the corresponding integrals are considerably more involved and do not admit the same simple analytical treatment. A detailed analysis of the massive case is, therefore, beyond the scope of the present work.

Having established the behavior of the purely geometric contribution in the massless case for $z=1$, we can now analyze the temperature-dependent contribution. Under the conditions obtained above, the corresponding quadratic equation reduces to
\begin{eqnarray}
    \Psi^2
    &=&   
    -\frac{3L^2e^{-\beta\sqrt{\mathcal{R}}}}{8\pi \mathcal{V}_E}<0\,.
\end{eqnarray}
Since all parameters entering this expression are positive, $\Psi^2$ is negative for any finite $\beta$, and therefore no real nontrivial solution exists. In the zero-temperature limit, $\beta\to\infty$, the exponential factor vanishes and the nontrivial solution continuously reduces to $\Psi=0$. 

Finally, in the massless case, $\Psi=0$ corresponds to the global minimum of the effective potential and is the unique stable vacuum configuration in the perturbative regime. In contrast, the massive case may exhibit additional local minima; however, a more detailed analysis of the corresponding integrals would be required to determine whether a global minimum exists.

Now, we calculate the Casimir energy density by evaluating the renormalized one-loop effective potential at $\Psi=0$, for an arbitrary anisotropic scaling factor. From $V_{\text{eff}}^{R}$ in \eqref{Veff2} and, taking into account the volume defined in \eqref{volume}, the Casimir energy density is given by
\begin{eqnarray}
\label{FullCasimirEnergy}
    \mathcal{E}_{C}
    &=&
    \left(a + \int_{-1/2}^{1/2}\int_{-1/2}^{1/2}\hat{f}(u_i)du_i\right)
    \left.V^{R}_{\text{eff}}(\Psi)\right|_{\Psi=0}
    \nonumber\\
    &=&
    \frac{\sin \left(\frac{\pi}{2} (z+2)\right)}{2\pi\eta_{z}(z+2)}
    \int_{m^{\frac{1}{z}}(\eta_{z}\mathcal{R})^{-\frac{1}{2}}}^\infty
    dx\left[
    \frac{\left((\eta_{z}\mathcal{R})^{z}x^{2z}
    - m^2\right)^{\frac{1}{z}+\frac{1}{2}} }
    {e^{2\pi x}-1}
    \right.
    \nonumber\\
    &&
    \left.
    \times\, _2F_1\left(-\frac{1}{2}-\frac{1}{z},
    1-\frac{1}{z};
    \frac{1}{2}-\frac{1}{z};
    \frac{m^2}{-(\eta_{z}\mathcal{R})^{z}x^{2z} + m^2}\right)\right]
    \nonumber\\
    &&
    - \frac{\mathcal{R}}
    {4\pi z \beta}
    \int_1^{\infty } \omega ^{\frac{1}{z}-1}
    e^{-\beta  \sqrt{\omega(\eta_{z}\mathcal{R})^{z} + m^2}} \, d\omega
    \,.
\end{eqnarray}
In the massless limit, this expression takes a simplified form
\begin{eqnarray}
\label{FullCasimirMassless}
    \mathcal{E}_{C}
    =
    \frac{\eta_{z}^{z/2}\,\mathcal{R}^{1+\frac{z}{2}}\sin \left(\frac{\pi}{2} (z+2)\right)}{2\pi(z+2)}
    \int_{0}^\infty
    \frac{x^{2 + z} }
    {e^{2\pi x}-1}dx
    - \frac{\mathcal{R}}
    {4\pi z \beta}
    \int_1^{\infty } \omega ^{\frac{1}{z}-1}
    e^{-\beta  \sqrt{\omega(\eta_{z}\mathcal{R})^{z}}} \, d\omega
    \,.
    \nonumber
    \\
\end{eqnarray}
Note that the integrals appearing in this result are defined by:
\begin{eqnarray}
    \int_{0}^\infty
    \frac{x^{2 + z} }
    {e^{2\pi x} - 1}\,dx 
    &=&
    \frac{\zeta_R\left(z+3\right) \Gamma (z+3)}{(2\pi)^{z+3}}\,,
    \\
    \int_1^{\infty } 
    \omega ^{\frac{1}{z}-1}
    e^{-\beta  \sqrt{\omega(\eta_{z}\mathcal{R})^{z}}} \, d\omega
    &=&
    2 E_{1 - \frac{2}{z}}\left(\beta  (\eta_z \mathcal{R} )^{z/2}\right)
    \,,
\end{eqnarray}
where $E_{1-\frac{2}{z}}$ denotes the generalized exponential integral, which is positive and convergent for positive arguments for all $z\geq1$.
Therefore, the Casimir energy density with anisotropic scaling factor dependence is
\begin{eqnarray}
\label{FullCasimirZ}
    \mathcal{E}_{C}
    =
    \frac{\eta_{z}^{z/2}\,\mathcal{R}^{1+\frac{z}{2}}\sin \left(\frac{\pi}{2} (z+2)\right)}{(2\pi)^{z+4}(z+2)}
    \,\zeta_R\left(z+3\right) \Gamma (z+3)
    - \frac{\mathcal{R}}
    {2\pi z \beta}
    E_{1 - \frac{2}{z}}\left(\beta  (\eta_z \mathcal{R} )^{z/2}\right).
\end{eqnarray}
As mentioned above, in the purely geometric term, only odd values of the anisotropic factor $z$ contribute to the energy density (this has already been demonstrated in the literature \cite{Cheng:2022mwd,Borquez:2023ajx}). In the case of the thermal term, any integer value of $z$ can be considered since the corresponding integral is positive, convergent and provides a nonvanishing contribution. However, this contribution exhibits a strong dependence on the parameter $\xi$. 

We are particularly interested in how the energy density depends on the roughness of the plates. A perturbative expansion in terms of roughness within the term $\mathcal{R}$ would yield the expression we are seeking. To better visualize the expansion, we set $z=1$. Using the perturbative expansion of $\mathcal{M}(u_{i})$ defined in \eqref{M}, the equation \eqref{FullCasimirZ} is systematically expanded in powers of the small parameter $\hat{f}/a$ contained in $\mathcal{R}$. The resulting Casimir energy is\footnote{For simplicity of notation, we have reduced the integral expression of the roughness in each term as follows: $\int_{-1/2}^{1/2}\int_{-1/2}^{1/2}du_{i}\rightarrow\iint du_{i}$.}
\begin{eqnarray}
    \mathcal{E}_{C} 
    &=&
    - \frac{\pi^{2}}{1440\,a^{3}}
    -\frac{1}{2a\beta^{2}}\left(1+\frac{a}{\pi\beta}\right)e^{-\frac{\pi\beta}{a}}
    + \frac{\pi}{2}\left(
    \frac{\pi}{240\,a^{4}}
    - \frac{e^{-\frac{\pi\beta}{a}}}{a^{3}\beta}
    \right)
    \iint\hat{f}(u_{i})du_{i}
    \nonumber\\
    &&
    -
    \frac{\pi}{4}\left[
    \frac{\pi}{60\,a^{5}}
    - \left(
     1
    - \frac{\pi\beta}{a}
    \right)\frac{e^{-\frac{\pi\beta}{a}}}{a^{4}\beta}
    \right]
    \iint\hat{f}^{2}(u_{i})du_{i}
    \nonumber\\
    &&
    + \frac{\pi}{2 a^{4}\beta}
    e^{-\frac{\pi\beta}{a}}\left(
    \iint\hat{f}(u_{i})du_{i}
    \right)^{2}
    + \mathcal{O}(\hat{f}^{3})
    \,.
\end{eqnarray}
The first term corresponds to the standard zero-temperature Casimir energy for flat plates, while the second term represents the leading thermal correction, exponentially suppressed in the low-temperature regime. The remaining terms describe roughness corrections, including both temperature-independent contributions and mixed roughness-temperature effects. In the limit of vanishing roughness and zero temperature, our result reduces to the standard Casimir energy density for a massless scalar field subject to Dirichlet boundary conditions, thereby providing a consistency check of our calculation \cite{Toms:1979ij,Junior:2023feu,Borquez:2026nos}.

Now, we analyze the topological mass, defined as the curvature of the renormalized effective potential at the stable vacuum $\Psi=0$. In the present setup, it is given by
\begin{eqnarray}
\label{Tmass}
    m^{2}_{T}
    &=&
    m^2
    +
    \frac{g\beta L^{2}}{\mathcal{V}_{E}}
    \left\{
    \frac{\sin \left(\frac{\pi}{2} (z+2)\right)}{4\pi z\eta_{z}m^{2}}
    \int_{m^{\frac{1}{z}}\left(\eta_{z}\mathcal{R}\right)^{-\frac{1}{2}}}^\infty
    dx\,
    \frac{\left(\left(\eta_z  \mathcal{R}\right)^z x^{2 z}
    - m^2\right)^{\frac{1}{z}+\frac{1}{2}}}{e^{2\pi x}-1} 
    \right.
    \nonumber\\
    &&
    \times
    \left[
    _2F_1\left(
    1 - \frac{1}{z}, - \frac{1}{2 } - \frac{1}{z};
    \frac{1}{2} - \frac{1}{z};
    \frac{m^2}{m^2 - (\eta_z \mathcal{R})^zx^{2 z}}
    \right)
    - \left(\frac{\left(\eta_z \mathcal{R}\right)^z
    x^{2z}}
    {\left(\eta_z \mathcal{R}\right)^zx^{2z}
    - m^2}\right)^{\frac{1}{z}}
    \right]
    \nonumber\\
    &&
    \left.
    + \frac{\mathcal{R}}{8\pi z }
    \int_1^{\infty }
    \omega ^{\frac{1}{z}-1}
    \frac{e^{-\beta  \sqrt{\omega (\eta_z \mathcal{R})^z + m^{2}}}}
    {\sqrt{\omega (\eta_z  \mathcal{R})^z + m^{2}}}
    \, d\omega
    \right\}
    \,.
\end{eqnarray}
The correction of the topological mass is entirely proportional to the coupling $g$ and therefore vanishes in the free-field limit $g\to0$. Thus, the mass shift arises solely from the quartic self-interaction, with its dependence on anisotropic scaling, boundary conditions, plate geometry, and temperature encoded in the corresponding terms of \eqref{Tmass}. For even values of $z$, the mass contribution arises exclusively from the thermal integral. Since this integral is positive, the sign of the coupling $g$ determines the sign of the corresponding contribution, and there exists a range of $g$ for which the topological mass remains positive. In the massless case, the positivity of the thermal integral restricts the result to positive values of $g$. For odd values of $z$, the topological mass is determined by both contributions, the purely geometric sector and the temperature sector. In this case, the original mass provides an intrinsic scale that regulates the infrared behavior and ensures that the resulting expression remains well defined. The massless case can be evaluated without any impediment,
\begin{eqnarray}
    \label{Tmassless}
    m^{2}_{T}
    =
    \frac{g\beta L^{2}\mathcal{R}^{1-\frac{z}{2}}}{4\pi\eta_{z}^{z/2}\mathcal{V}_{E}}
    \left[
    \frac{\sin \left(\frac{\pi}{2} (z+2)\right)}{ (z-2)}
    \int_{0}^\infty
    dx\,
    \frac{x^{2-z}}{e^{2\pi x}-1} 
    + \frac{1}{2 z }
    \int_1^{\infty } 
    \omega ^{\frac{1}{z}-\frac{3}{2}}
    e^{-\beta  \sqrt{\omega  \left(\eta_z \mathcal{R}\right)^z}}
    \, d\omega
    \right]
    \,.
    \nonumber\\
\end{eqnarray}
This expression can be consistently evaluated only within the range of $z$ for which the purely geometric contribution remains finite. This occurs when the integration converges, that is, $z<2$. Since $z$ is restricted to positive integer values, the only case in which the massless limit is finite is $z=1$. For odd integer values $z>1$, the integration of the first term develops an infrared divergence as $x\to 0$; this divergence is not induced by the boundary roughness, but arises from the absence of an intrinsic mass scale; and the massless theory becomes increasingly sensitive to infrared modes as the anisotropic scaling exponent $z$ increases. For the case $z=1$, as for the Casimir energy density, it is possible to perform a perturbative expansion in powers of $\hat{f}/a$. Therefore, the topological mass is given by
\begin{eqnarray}
\label{Tmasslessfinal}
    m_{T}^{2}
    &=&
    \frac{g}{96a^{2}}
    + \frac{g}{4\pi\beta a}e^{-\frac{\pi\beta}{a}}
    + \frac{g}{4}\left[
    - \frac{1}{12a^{3}}
    + \left(
    \frac{1}{a^{3}}
    - \frac{1}{\pi\beta a^{2}}
    \right)
    e^{-\frac{\pi\beta}{a}}
    \right]
    \iint\hat{f}(u_{i})du_{i}
    \nonumber\\
    &&
    + \frac{g}{4}\left[
    \frac{1}{12a^{4}}
    + \left(
    \frac{1}{\pi\beta a^{3}}
    + \frac{(\pi\beta - 2a)}{2a^{5}}
    \right)
    e^{-\frac{\pi\beta}{a}}
    \right]
    \iint\hat{f}^{2}(u_{i})du_{i}
    \nonumber\\
    &&
    + \frac{g}{4}
    \left(
    \frac{1}{24a^{4}}
    -\frac{e^{-\frac{\pi\beta}{a}}}{a^{4}}
    \right)
    \left(
    \iint\hat{f}(u_{i})du_{i}
    \right)^{2}
    + \mathcal{O}\left(\hat{f}^{3}\right)
    \,.
\end{eqnarray}
Within the approximations considered, namely, the perturbative treatment of the boundary roughness and the low-temperature regime, no stability problems arise for $g>0$. This indicates that the corresponding configuration remains stable within the validity regime of our analysis. In the limit of vanishing roughness, it is possible to recover the known result reported in \cite{Borquez:2026nos}, and evaluating the case $\beta\rightarrow\infty$, we see that the results coincide with those obtained in \cite{Toms:1979ij,Junior:2023feu,Farias:2024uzf}.


\subsection{Two-loop correction}
Having obtained the one-loop contribution to the Casimir energy, we now calculate the leading contribution from the quartic self-interaction at two-loop order. Since the one-loop Casimir energy is independent of the coupling $g$, the first interaction-dependent correction arises at two-loop order. Therefore, we evaluate the two-loop Casimir energy for a vanishing background field $\Psi=0$. At this order, the relevant contribution is given by the one-particle-irreducible vacuum diagram with a single quartic vertex, that is, the standard \emph{figure-eight} diagram \cite{Toms:1979ij}. We evaluate this contribution within the $\zeta$-function regularization framework introduced above. The resulting two-loop Casimir energy is expressed in terms of the spectral sum associated with the anisotropic quadratic operator, retaining its dependence on plate separation, geometry, and temperature. For the anisotropic theory considered here, this contribution remains finite at $s=1$. Therefore, the two-loop effective potential is given by
\begin{eqnarray}
\label{two_loops_mass}
    V^{(2)}_{E}(0)
    &=&
    \left.
    \frac{g}{8}
    \left(\frac{\zeta_{\hat{Q}}\left(1\right)}{\mathcal{V}_E}\right)^2
    \right|_{\Psi=0}
    \nonumber\\
    &=&
    \frac{g}{8}\left\{
    \frac{\beta\pi}{z\eta_{z}\mathcal{V}_E}
    \left(
    \frac{L}{2\pi}
    \right)^2
    \left[
    \frac{2z\sin\left(\frac{\pi z}{2}\right)}{2-z}
    \int_{m^{\frac{1}{z}}(\eta_{z}\mathcal{R})^{-\frac{1}{2}}}^\infty
    \frac{\left((\eta_{z}\mathcal{R})^{z}x^{2z}
    - m^2\right)^{-\frac{1}{2}+\frac{1}{z}}}
    {e^{2\pi x}-1}
    \right.
    \right.
    \nonumber\\
    &&
    \times
    {}_2F_{1}\left(
    \frac{1}{2} - \frac{1}{z},
    1  - \frac{1}{z};
    \frac{3}{2} - \frac{1}{z};
    \frac{m^2}{-(\eta_{z}\mathcal{R})^{z}x^{2z}+m^2}\right)\,dx
    \nonumber\\
    &&
    \left.
    \left.
    + \eta_{z}\mathcal{R}
    \int_1^\infty d\omega\,
    \omega^{\frac{1}{z}-1}
    \left(\omega(\eta_{z}\mathcal{R})^{z} + m^{2}\right)^{-\frac{1}{2}}
    e^{-\beta\sqrt{\omega(\eta_{z}\mathcal{R})^{z} + m^{2}}}
    \right]
    \right\}^2\,.
\end{eqnarray}
 The two-loop Casimir energy density contribution is determined by $\mathcal{E}^{(2)}_C=\left(a+\iint\hat{f}\right)V^{(2)}_{E}(0)$.
 In the massless case, we have
\begin{eqnarray} 
\label{two_loops-massless} 
    \mathcal{E}^{(2)}_C 
    &=& 
    \frac{g\mathcal{R}^{2-z}}{128\pi^{2}z^{2}\eta_{z}^{z}\left(a + \iint\hat{f}(u_{i})du_{i}\right)}
    \left( \frac{2z\sin\left(\frac{\pi z}{2}\right)}{2-z} 
    \int_{0}^\infty
    \frac{x^{2-z}}{e^{2\pi x}-1}\,dx 
    \right.
    \nonumber\\ 
    && 
    \left. 
    + \int_1^\infty d\omega\, 
    \omega^{\frac{1}{z}-\frac{3}{2}} 
    e^{-\beta\sqrt{\omega(\eta_{z}\mathcal{R})^{z}}} \right)^2\,. 
\end{eqnarray}
The massless case can be evaluated only within the range of $z$ for which the purely geometric contribution remains finite. The first integral  has the same infrared divergence origin as discussed above in the analysis of the topological mass. Accordingly, the same infrared restriction applies to the present two-loop contribution. In particular, for $z=1$, after performing the integrations of both contributions, the two-loop correction to the vacuum energy density, perturbatively evaluated with respect to the surface roughness in powers of $\hat{f}/a$, as was done for the one-loop case, is given by
\begin{eqnarray}
    \mathcal{E}^{(2)}_C
    &=&
    g\left\{
    \frac{1}{18432a^{3}}
    + \frac{e^{-\frac{\pi\beta}{a}}}{384\pi a^{2}\beta}
    + \left(
    - \frac{1}{6144a^{4}}
    + \frac{(\pi\beta - 2a)}{384\pi a^{4}\beta}
    e^{-\frac{\pi\beta}{a}}
    \right)
    \iint\hat{f}(u_{i})du_{i}
    \right.
    \nonumber\\
    &&
    +
    \left[
    \frac{1}{9216a^{5}}
    + \frac{1}{144a^{4}}
    + \left(
    \frac{(\pi\beta - 2a)}{768a^{6}}
    + \frac{(1 + 128a)}{384\pi a^{4}\beta}
    \right)e^{-\frac{\pi\beta}{a}}
    \right]
    \iint\hat{f}^{2}(u_{i})du_{i}
    \nonumber\\
    &&
    \left.
    + \left[
    \frac{1}{72a^{4}}
    + \frac{1}{18432a^{5}}
    + \left(
    - \frac{(1+ 128a)}{384a^{5}}
    + \frac{1}{3\pi a^{3}\beta}
    \right)e^{-\frac{\pi\beta}{a}}
    \right]
    \left(\iint\hat{f}(u_{i})du_{i}
    \right)^{2}
    + \mathcal{O}\left(\hat{f}^{3}\right)
    \right\}
    \,,
    \nonumber\\
\end{eqnarray}
where we have suppressed higher-order exponential terms involving temperature. The surface roughness introduces additional corrections to the two-loop Casimir energy density through the terms linear and quadratic in the roughness profile. Upon specifying the profile $\hat{f}$, the remaining integrals can be explicitly evaluated, yielding the corresponding two-loop Casimir energy density. In the limit $\hat{f}=0$, the result reduces to the known expression for flat boundaries \cite{Toms:1979ij,Farias:2024uzf}. These results show how the surface roughness and thermal effects modify the two-loop Casimir energy in the presence of quartic self-interaction.


\section{Conclusions}

In this work, we have investigated the Casimir effect for a self-interacting real scalar field with quartic coupling in a $(3+1)$-dimensional Ho\v{r}ava-Lifshitz-type theory, in the presence of rough Dirichlet boundaries and at low temperature. Using the effective-action formalism and a generalized spectral $\zeta$-function approach, we have derived the renormalized one-loop effective potential and obtained the corresponding Casimir energy and topological mass. Also, we have calculated the leading interaction-dependent correction to the Casimir energy at two-loop order. This framework has allowed us to study the combined effects of anisotropic scaling, boundary roughness, temperature, and self-interaction. Within the renormalization prescription adopted here, after imposing the renormalization conditions, all contributions depend on the plate separation $a$. Consequently, in the limit $a\to\infty$, all contributions vanish, leaving no residual counterterm contribution, and the resulting one-loop effective potential remains finite.

At one-loop order, the vacuum stability is analyzed perturbatively in $g$. For even $z$, the purely geometric contribution vanishes, while the thermal contribution remains finite and positive. In this case, the nontrivial solutions do not correspond to stable vacua, including in the massless case. For odd $z$, the purely geometric contribution modifies the stability of the vacuum, and stable nontrivial solutions may occur for specific values of the physical parameters of the theory. However, a detailed analysis of this is hindered by the complexity of the geometric integral. On the other hand, in the massless case, the anisotropic parameter is restricted to $z=1$, allowing the geometric part to vanish in this limit. Regarding temperature, the application of these conditions yields a positive and convergent contribution, leading to the conclusion that no real solution exists for $\Psi$. Therefore, for any $z$, $\Psi=0$ is the unique stable vacuum in the massless case within the perturbative regime.

The one-loop topological mass receives a contribution proportional to the coupling $g$, whereas the Casimir energy is independent of $g$ at this order. The result obtained for the Casimir energy indicates that it is fully convergent for $z \geq 1$ and strongly depends on the anisotropic scaling. For even values of $z$, as previously mentioned, the purely geometric sector vanishes, allowing only the temperature sector to contribute. The topological mass contains both geometric and thermal contributions, with the dependence on the boundary roughness incorporated through the perturbative expansion of the rough geometry. For even $z$, the geometric contribution vanishes, so that the topological mass is entirely determined by the finite thermal contribution. For odd $z$, both geometric and thermal contributions are present. In the massless limit, the purely geometric contribution is finite only for $z=1$ among the odd integer values of $z$; for odd $z>1$, it exhibits a power-law infrared divergence arising from the absence of an intrinsic mass scale, rather than from the boundary roughness. The thermal contribution remains finite and therefore does not alter the infrared restriction associated with the geometric part. Thus, the temperature provides a finite correction to the topological mass. In the simultaneous smooth-boundary, zero-temperature, and $z=1$ limits, the result reduces to the standard relativistic Dirichlet result.

At two-loop order, the leading interaction-dependent correction to the Casimir energy is obtained from the \emph{figure-eight} vacuum diagram containing a single quartic vertex. This contribution is proportional to $g$ and constitutes the first self-interaction-dependent correction to the Casimir energy. Both temperature and boundary roughness contribute to this correction, with the roughness effects incorporated through the perturbative expansion of the boundary profile. The infrared convergence of the massless contribution is governed by the same condition discussed above for the topological mass, so no additional restriction arises at this order.

Overall, our results show that anisotropic scaling, boundary roughness, temperature, and self-interaction modify the Casimir interaction in distinct but interconnected ways. The one-loop analysis determines the vacuum structure, topological mass, and Casimir energy, whereas the two-loop calculation provides the leading self-interaction-dependent correction to the Casimir energy. Within the perturbative regime considered, the massless theory admits $\Psi=0$ as its unique stable vacuum. A systematic investigation of the massive regime and of higher-order interaction corrections constitutes a natural extension of the present work.



\end{document}